# Back-end and Flexible Substrate Compatible Analog Ferroelectric Field Effect Transistors for Accurate Online Training in Deep Neural Network Accelerators

*Sayani Majumdar\* and Ioannis Zeimpekis*

Sayani Majumdar

VTT Technical Research Centre of Finland Ltd., P.O. Box 1000, FI‐02044, VTT, Espoo, Finland

E-mail: sayani.majumdar@vtt.fi

Ioannis Zeimpekis

Zepler Institute, Faculty of Engineering and Physical Sciences, University of Southampton, United Kingdom

Keywords: ferroelectric field effect transistor, analog memory, electronic synapses, online training, deep neural network

Online training of deep neural networks (DNN) can be significantly accelerated by performing in-situ vector matrix multiplication in a crossbar array of analog memories. However, training accuracies often suffer due to device non-idealities such as nonlinearity, asymmetry, limited bit precision and dynamic weight update range within constrained power budget. Here, we report a three-terminal Ferroelectric-Field-Effect-Transistor based on low thermal budget processes that can work efficiently as an analog synaptic transistor. Ferroelectric polymer P(VDF-TrFE) as the gate insulator and 2D semiconductor $MoS_2$ as the n-type semiconducting channel material makes them suitable for flexible and wearable substrate integration. The analog conductance of the FeFETs can be precisely manipulated by employing a ferroelectric-dielectric layer as the gate stack. The ferroelectric-only devices show excellent performance as digital non-volatile memory operating at $<\pm5V$ while the hybrid ferroelectric-dielectric devices show quasi-continuous resistive switching resulting from gradual ferroelectric domain rotation, important for their multibit operation. Analog conductance states of the hybrid devices allow linearity and symmetry of weight updates and produce a dynamic conductance range of 104 with >16 reproducible conducting states. Network training experiments of these FeFETs show >96% classification accuracy with MNIST handwritten datasets highlighting their potential for implementation in scaled DNN architectures.

## 1. Introduction

In recent years, deep neural networks (DNN) have shown promising performance in different cognitive tasks like image and speech recognition. [1,2] Typically, DNN training is performed through off-chip memory access and therefore the energy consumption and training time of DNNs are limited by the memory bandwidth which creates a bottleneck. It has been demonstrated previously that for a fully connected DNN, significant acceleration in training can be achieved by minimizing data movement between off-chip storage and processor. [3-5] This requires on-chip storage and processing of weights at the same node, with all nodes connected together in an array. Monolithic crossbar / pseudo crossbar arrays with analog non-volatile memories can store and update on-chip weights and therefore accelerate DNN training significantly.

Analog non-volatile memory elements with multiple distinguishable states are under intensive studies to enable integrated matrices of synaptic weights. [6,7] Resistive random-access memory (RRAM) and phase change memory (PCM) are potential candidates due to their small cell size ($4F^2$) and the technology readiness level. However, achieving reproducible analog states, linearity and symmetry in potentiation and depression characteristics, is a challenge in these devices. [8,9] Ultralow power non-volatile ferroelectric memories have emerged as an attractive solution in this respect. [10,11] However, material choice for analog ferroelectric memories are often limited by demanding process parameters that make them incompatible for CMOS back-end-of-line (BEOL) and flexible platform integration. In our recent work, we have shown that precise control over ferroelectric domain dynamics in polymer ferroelectric P(VDF-TrFE) can result in multiple distinguishable states in ferroelectric tunnel junctions (FTJs), [12] that lead to ultrafast operation, [13] conductance linearity and efficient DNN training. [14] However, there is a compromise between the maximum and minimum conductance ratio ($G_{max}/G_{min}$) and the linearity of the weight update in these devices. For higher efficiency in DNN online training in larger networks, $G_{max}/G_{min}$ ratio and linearity needs to improve.

In this context, three terminal Ferroelectric-Field Effect Transistors (FeFET) can provide a better solution when compared to RRAM and PCM based devices. [10, 15] A FeFET may act as a one-transistor memory where the gate insulator is composed of a ferroelectric thin film and information is stored in the ferroelectric layer as voltage-controlled polarization states. In these devices the dynamics of voltage-controlled partial polarization switching can be utilized to achieve analog synaptic features. FeFET based synaptic transistors have been demonstrated in recent times, however, precise control over their analog states, CMOS compatibility of the devices, considerations for monolithic integration, and scalability require exploration over vast range of enabling materials and potential designs.[10]

FeFET performance depends critically on channel semiconductor and gate ferroelectric properties. 2D semiconductors such as transition metal dichalcogenides (TMDs) have emerged as CMOS post-processing compatible ultrathin semiconductors with potential for device scalability, as their few-atom thick layers are devoid of surface dangling bonds. Layered TMDs can be ideal channel materials due to their robustness against short-channel effects down to a few atoms short gate lengths. [16, 17] Among TMDs, molybdenum disulfide ($MoS_2$) is one of the most promising materials due of its relatively high chemical stability and large bandgap (1.8–1.9 eV). Moreover, the excellent properties of $MoS_2$ as the transistor channel material such as large On/Off ratio, steep subthreshold slope (SS) and high *n*-type channel mobility provide a promising route towards high-density integration of memory arrays where a low operational power is highly desired.

Ferroelectric based memory devices ideally exhibit non-volatility, fast switching, a high On/Off ratio, and programmable multilevel conductance states provided certain conditions are fulfilled. [18] For instance, non-volatility requires proper screening of the bound polarization charges for the depolarizing field to be minimal. [19] In FeFETs, the gate stack often contains a dielectric on one side of the ferroelectric that makes the screening incomplete, increasing the depolarizing field and causing finite retention. [20] Additionally, charge trapping sites created at the ferroelectric-dielectric interface leads to faster breakdown of the devices over multiple switching cycles. [21] Therefore, careful selection of dielectric and ferroelectric materials is of vital importance. In FeFETs, large On/Off switching ratio depends on $180^0$ polarization reversal of the ferroelectric domains to achieve full accumulation and depletion states in the semiconductor channels. Furthermore, analog conductance for neuromorphic training requires voltage pulse-controlled precise partial rotation of ferroelectric domains. Transistors with integrated ferroelectrics in their gate stack can retain the programmed states without any applied gate bias and have shown varying degrees of retention in previous reports, i.e. non-volatile to volatile to programmable synaptic weight retention and analog conductance based on gate stack composition, geometry, semiconductor carrier density, channel area etc. [22, 23] In spite of these attractive features, ferroelectric technologies are yet to achieve commercial uptake due to the demanding fabrication methodologies and their incompatibility with low thermal budget processes.

The two most suitable candidates for CMOS back-end compatible ferroelectric are doped $HfO_2$ films and the organic ferroelectric P(VDF-TrFE). Their main advantages and challenges have been discussed in detail in a recent review. [18] Interest in organic ferroelectric materials [24, 25] is mainly motivated by the low processing temperature and structural flexibility of the ferroelectric polymers that allows device fabrication on virtually any platform including CMOS chips, plastic or paper foils and even fabrics, enabling their application in flexible and wearable smart devices. [26, 27] The most reported organic ferroelectric materials are poly(vinylidenefluoride) (PVDF) and its random copolymers with trifluoroethylene (P(VDF-TrFE)). Solution-processed organic FeFETs based on PVDF and P(VDF-TrFE) in combination with various semiconducting polymers and metal-oxides have been reported, [18, 28, 29] including a reconfigurable 250 kbit memory array. [30] In all cases, the operation voltage of the transistors is large, typically more than 20 V, which makes them unsuitable for integration with CMOS readout platform or flexible smart systems. In CMOS, the need for higher driving voltage necessitates larger MOSFETs that restrict the integration density and energy-efficiency. For flexible, wearable and edge devices, larger operating voltage means lower battery lifetime which is undesirable for most use cases.

In most previous reports with P(VDF-TrFE) based FeFETs, a P(VDF-TrFE) film thickness of 100 nm or more is used to reduce the gate leakage. However, thick ferroelectric layers have the obvious effect of higher operating voltages. In recent times, several P(VDF-TrFE) based FTJ memories were reported in literature. [31, 32] This proves high quality ultrathin films of P(VDF-TrFE) can be made using Langmuir-Blodgett or spin-coating techniques, enabling devices where polarization switching takes place below 3V. Depending on the film morphology and crystallinity, switching can be achieved at the sub-nanosecond timescale and provide control over multiple analog conductance states. [13] In another report, it was shown that thin ferroelectric layers of P(VDF-TrFE-CTFE) in conjunction with an ultrathin high-$k$ oxide layer of $AlO_x$ can reduce the operating voltage to ± 4V not compromising the gate leakage current, that is essential for long time stability of the memory elements.[33]

Another potential route for reducing the operational voltage is using a semiconductor that can provide a steep subthreshold slope. International Technology Roadmap for Semiconductors (ITRS) [34] predicts that $MoS_2$ is a potential semiconductor material for integration into nanometer scale structures due to outstanding electrical properties exhibited by one or few monolayer thick films. [14,15] $MoS_2$ or other 2D semiconductor-based circuits can provide a solution to this high operating voltage problem of the FeFETs due to an indirect bandgap of 1.2 eV of few layer $MoS_2$, that resulted in transistors with a high On/Off current ratio of ~$10^6$–$10^8$ under a low operating voltage. Also, $MoS_2$ MOSFETs exhibited a very low SS value of 74 mV/decade, benefiting from the absence of dangling bonds in the $MoS_2$ layers. [16, 17] In most of the earlier studies, exfoliated flakes of $MoS_2$ were used to fabricate $MoS_2$ based FeFETs. [35] Although promising reports of large On/Off switching, steep SS and considerable memory window (MW) has been reported in these studies, a necessary requirement for circuit integration of these devices, i.e. large area wafer scale uniformity of the $MoS_2$ film is missing in these works. [35, 36]. In our recent work, we reported one-transistor (1T) FeFET memory cell fabricated using large area ALD grown $MoS_2$ films [37] with polymeric ferroelectric P(VDF-TrFE) as the gate layer. In our current work, we demonstrate FeFET devices where ferroelectric polarization can be tuned controllably to produce both full and partial polarization switching under operating voltage below ±4V which makes them suitable for low-power operating memory elements. High-quality wafer scale $MoS_2$ as the channel semiconductor enables the scalability of the technology. By tuning the gate stack composition, we demonstrate both digital or analog modulation of the channel conductance, highlighting the high-performance of the devices as programmable synaptic elements in neuromorphic architectures. The ferroelectric-only devices show digital memory characteristics with On/Off ratio of $10^6$, memory window of 3V and non-volatile data retention operating at <±5V while devices with a composite ferroelectric (FE) – dielectric (DE) gate stack showed more than 16 distinguishable conductance states with good linearity and symmetry, operating below 4V. DNN online training simulation using analog conductance states from these devices showed > 96% accuracy, making them one of the most promising building blocks for online DNN training accelerators. In addition to the excellent performance, the mechanical flexibility of the ultrathin semiconductor and polymer ferroelectric makes our FeFETs suitable for integration in wearable cognitive devices.

## 2. Results & Discussion

### 2.1. FeFET operation principle and characteristics

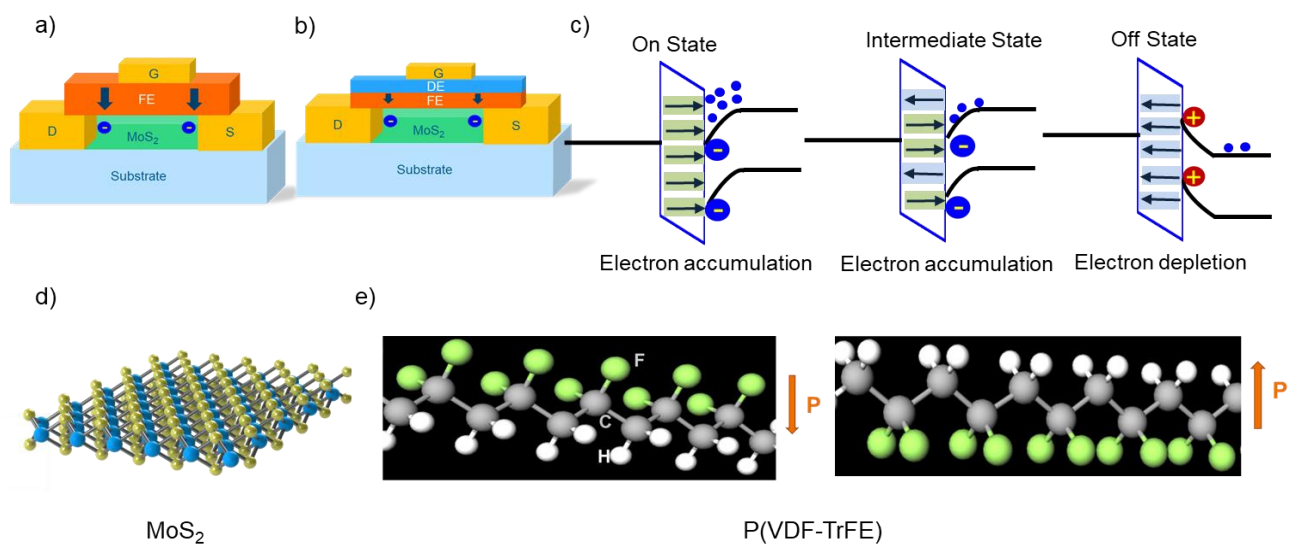

**Figure 1.** Device schematics of two types of FeFETs with (a) only ferroelectric P(VDF-TrFE) and (b) combined ferroelectric and dielectric (P(VDF-TrFE) and $Al_2O_3$) as the gate layer. (c) Operation principles of the FeFETs describing the digital and analog switching of the two structures. Molecular structure of (d) 2D semiconductor $MoS_2$ and (e) P(VDF-TrFE) in down and up polarized states.

**Figure 1** shows the schematic diagram of the two FeFET architectures under study, the operating principles of the FeFET and molecular structure of the $MoS_2$ and P(VDF-TrFE) molecules. The details of the fabrication process flow are given in supplementary information **Figure S1**. For a typical FeFET, when a gate voltage $V_G$ of certain polarity is applied, the ferroelectric polarization direction change at the gate stack that creates an accumulation or depletion of the charge carriers at the semiconductor channel of the field-effect-transistors (FETs). For an *n*-type semiconductor channel, when the positive gate bias exceeds the coercive voltage, $V_c$ of the ferroelectric, the P(VDF-TrFE) polarization experiences an abrupt change and polarization direction points towards the $MoS_2$ channel. Since $MoS_2$ is an *n*-type semiconductor, the channel supplies electrons for compensation and stabilization of the positive polarization bound charges. As the ferroelectric polarization is much larger than the dielectric charge density, the current increases abruptly at $V_c$ and the transistor reaches the On-state. Upon scanning $V_G$ backwards to zero, the ferroelectric polarization remains in the same state and the current remains high. Upon further increase of $V_G$ to the negative value, the polarization reverses as the negative coercive voltage, $-V_c$, is exceeded. The reverse polarization of the ferroelectric gate causes negative surface charges to accumulate near the channel. Since the semiconductor is *n*-type, it cannot provide enough holes as compensating charges. Therefore, the channel reaches the depletion mode and as a consequence, the current drops and the transistor switches back to its Off state. In a *p*-type ferroelectric transistor, a reverse switching behavior is observed, i.e., the ferroelectric is negatively polarized in the On state and positively polarized in the Off state.

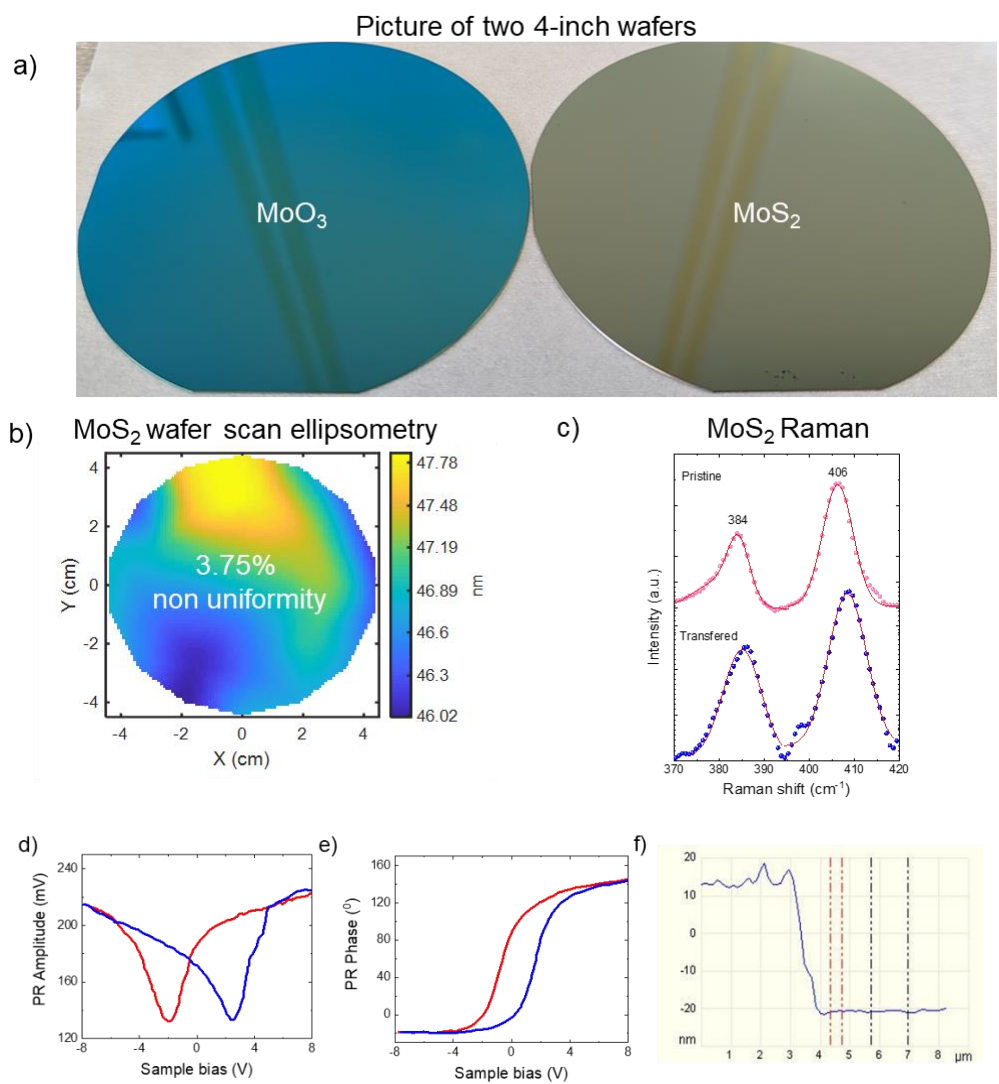

**Figure 2.** (a) Picture of 4-inch wafer sized $MoO_3$ and $MoS_2$ films. (b) Wafer-scale $MoS_2$ thickness profile from Ellipsometry measurement. (c) Raman spectroscopy of the pristine (top) and transferred (bottom) $MoS_2$ films. Piezoresponse (d) amplitude and (e) phase rotation under sweeping electric field for 30 nm P(VDF-TrFE) film on ITO substrate measured using piezo-force microscopy (PFM). (f) Atomic Force Microscope thickness profile from a P(VDF-TrFE) film on ITO substrate grown with identical deposition and annealing parameters and P(VDF-TrFE) solution strength.

**Figure 2** shows the basic material properties of the 4-inch wafer scale $MoS_2$ and 30 nm ferroelectric film. **Figure 2 (a)** shows a picture of two 4-inch wafers with large area continuous films of $MoO_3$ and $MoS_2$. The process we have followed is described in our previous work [37] and relies on ALD to create a uniform large $MoS_2$ film. ALD is used to cover the wafer with a $MoO_3$ film that serves as a template precursor. Following that an anneal in an $H_2S$ environment converts the $MoO_3$ to $MoS_2$. Finally, a high temperature anneal at 970°C crystallizes the film.

The advantage of the process is that it can achieve tunable, uniform large area films with high electrical performance. This is achieved by the decoupling of the layer number count of the film from its stoichiometry and crystallinity. **2(b)** shows the thickness fit from ellipsometric data for a 4-inch wafer. This deposition was performed with 300 ALD cycles as a uniformity comparison. The worst-case non-uniformity of the $MoS_2$ film ($T_{max}$-$T_{min}$/$T_{ave}$) was found to be 3.75% which is equivalent to our previous work for $MoO_3$ at the same wafer area.[37] This result verifies that the film thickness uniformity is defined during the $MoO_3$ ALD and is not detrimentally affected during the conversion to $MoS_2$. This is a significant advantage of the process as it shows it is scalable to arbitrary size. The films used to make the transistors were created using 15 ALD cycles and were measured at 1.48 nm. **Figure 2 (c)** shows Raman spectroscopy of the as-grown and transferred 15 cycles $MoS_2$ film. The characteristic $E_{2g}^1$ and $A_{1g}$ peaks of the $MoS_2$ corresponds to the in-plane vibration of two S atoms with respect to the Mo atoms and the out-of-plane vibration of S atoms in the opposite direction, respectively. The $MoS_2$ film had the $E_{2g}$ peak at 384 and the $A_{1g}$ at 406 cm$^{-1}$. The observed difference between $E_{2g}^1$ and $A_{1g}$ is 22 cm$^{-1}$, which is in good agreement with the reported mode difference of the $MoS_2$ for 2 monolayer thickness. To confirm the quality of the $MoS_2$ after a carrier-film assisted transfer process, Raman measurements were done on the transferred film again. For the transferred film on PI substrate, the peaks were at 385 and 408 cm$^{-1}$, confirming a damage free transfer of the $MoS_2$. **Figure 2 (d)** and **(e)** shows piezo-force microscopy (PFM) results of the 30 nm P(VDF-TrFE) films made on a conducting Indium-tin-oxide (ITO) substrate. The ferroelectric switching properties, demonstrated by the butterfly-shaped amplitude curves and near-180° phase contrast (**Figure 2(d)** and **(e)** respectively) confirms ferroelectricity of the films. The coercive voltage ($V_C$) of approximately 1.5 V, indicate coercive fields of the films to be ~50 MV/m which is in good agreement with other published results.[38, 39] The switching voltages vary slightly based on the bottom electrodes due to the difference in film morphology. It has been shown previously that the polarization and coercive voltages of P(VDF-TrFE) films vary with the crystallinity, size and orientation of grains.[32] **Figure 2(f)** shows the thickness of the P(VDF-TrFE) films to be ~30 nm measured with atomic force microscope (AFM) on the prepared step edges of the samples.

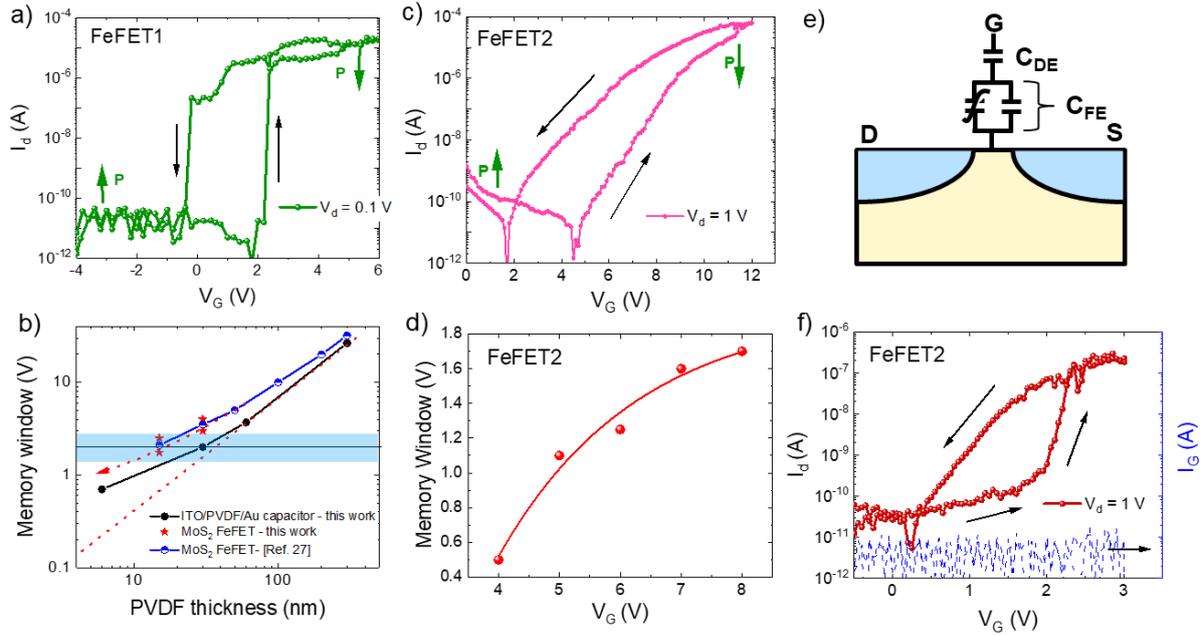

**Figure 3.** (a) Transfer characteristics of a typical FeFET1 device with 30 nm P(VDF-TrFE) as the gate layer and (b) Scaling of memory window as a function of ferroelectric thickness in ferroelectric MFM capacitors and FeFETs. (c) Transfer characteristics of a typical FeFET2 device with 10 nm P(VDF-TrFE) and 3 nm $Al_2O_3$ as the gate layer where addition of a dielectric layer in the gate stack clearly modifies the voltage drop across the ferroelectric layer making the polarization reversal more gradual. (d) Memory window dependance to different gate voltage ranges demonstrating partial polarization reversal in the P(VDF-TrFE) layer. (e) Voltage divider circuit model of a FeFET with a composite dielectric-ferroelectric gate stack and (f) partially open counterclockwise hysteretic $I_d$-$V_G$ loop of FeFET2 with $L$=10 µm, $W$=20 µm channel under gate sweeping range -1V to +3V.

**Figure 3(a)** shows typical transfer characteristics for FeFET1 (device with 30 nm P(VDF-TrFE) as the gate layer without $Al_2O_3$) when the top gate bias sweeping scheme is utilized. The voltage sweeping direction is from negative to positive and then back to negative thus allowing an Off to On transition and back to Off state again. The devices act as typical FeFETs with *n*-channel accumulation mode exhibiting a counter-clockwise hysteresis behaviour. For the FeFET1 with 30 nm P(VDF-TrFE) as the gate insulator layer, the threshold voltage for Off to On transition, $V_{th1}$, is positive while that for On to Off transition $V_{th2}$ is found to be negative, thus exhibiting a finite memory window (MW) of 3V and non-volatile retention of the programmed state, once the programming voltage is withdrawn. The On-state drain-source current ($I_d$) of ~$10^{-5}$ A was observed in the FeFET1 devices leading to a device On/Off ratio of $10^6$. The On/Off ratio at a reading voltage of $V_G$ = 0 V is measured to be ~ $10^4$.

Multiple FeFET1 devices were measured and, in all cases, an abrupt resistive switching was observed, corresponding to FE polarization switching field in the 30 nm FE layer. A MW of 3V is measured for FeFET1 when full polarization switching of the 30 nm ferroelectric layer is achieved with a sweeping voltage of ±5V, which is in accordance with the PFM and capacitance data (shown in supplementary information, **figure S3**). To obtain this MW and On/Off ratio, a sweeping voltage of ±5V is needed. To obtain a lower operation voltage, one possible route is ferroelectric thickness scaling. It has been shown in previous works that for thicker P(VDF-TrFE) layers (thickness >100 nm), the FeFET memory window scales almost linearly with gate insulator thickness. [27] To estimate the memory window with decreasing thickness of P(VDF-TrFE), we prepared a series of metal-ferroelectric-metal (MFM) capacitors with ITO as bottom electrode and Au as top electrode. Measurement of capacitance - voltage (*C-V*) curve in these devices (**figure S3**) show how thickness scaling at the gate stack can influence the FeFET memory window. **Figure 3(b)** shows data points obtained from our capacitance measurements, FeFET measurements and data points obtained from ref. [27]. All experimentally measured data and their extrapolation indicate that at the lower thickness range, below 50 nm, the scaling law deviates from a linear behaviour. Therefore, to achieve a full polarization switching below ±1.8 V, a P(VDF-TrFE) thickness of ~10 nm or less would be needed. Polarization switching below ±1.8 V is crucial as this is the limit for CMOS driving transistors in 180 nm technology node. However, a thin polymer layer of 10 nm has a high probability of current leakage through the gate. To minimize the gate leakage, a thin AlOx layer on one or both sides of the ferroelectric layer is an effective route [26] and therefore, we prepared the following batch of FeFETs, referred to as FeFET2, where a composite dielectric-ferroelectric barrier is used. A 3 nm ALD grown $Al_2O_3$ layer on top of 10 nm P(VDF-TrFE) was used in the FeFET2 devices. **Figure 3(c)** shows the transfer characteristics of the FeFET2 device showing a much more gradual switching from Off to On state indicating gradual polarization switching of the FE layer. For different $V_G$ sweeping ranges, the MW measurements showed a significant drop in MW when the devices are operated below 5V (**Figure 3(d)**). The hysteresis characteristics obtained in these devices can be understood following the principle of a ferroelectric capacitor in series with a dielectric capacitor in the gate stack. One ferroelectric capacitor is composed of one nonlinear hysteretic capacitor in parallel with a linear dielectric capacitor that works in series with a dielectric capacitor to act as a voltage divider in the gate stack as per the equation 1 and, as shown in the schematics in **Figure 3(e)**:

$$C_{Stack} = (C_{lin} + C_{nonlin})\ \text{II}\ C_{DE} = C_{FE}\ \text{II}\ C_{DE} \qquad (1)$$

where $C_{FE}$ and $C_{DE}$ are ferroelectric and dielectric capacitances, respectively and both capacitors are assumed to have the same area. This voltage division restricts the possibility of linear scaling of voltage across the ferroelectric as the gate voltage gets divided in different components. The electric field across the ferroelectric and the dielectric layer can be expressed as,

$$E_{DE} = \frac{\frac{V_G}{\frac{\varepsilon_{DE} \cdot d_{FE}}{\varepsilon_{FE} \cdot d_{DE}}+1}}{d_{DE}}, E_{FE} = \frac{\frac{V_G}{\frac{\varepsilon_{FE} \cdot d_{DE}}{\varepsilon_{DE} \cdot d_{FE}}+1}}{d_{FE}} \quad (2)$$

where $V_G$ is the applied gate voltage, $\varepsilon_{DE}$ and $\varepsilon_{FE}$ are the effective dielectric and ferroelectric permittivity and $d_{DE}$ and $d_{FE}$ are the oxide and ferroelectric thickness respectively. The dielectric constant ($\varepsilon$) for ALD grown $Al_2O_3$ varies between 7 to 8.4. However, with decreasing thickness, $\varepsilon$ starts to decrease and for 3 nm, $Al_2O_3$, the $\varepsilon_{DE}$ is ~ 4. [40] For P(VDF-TrFE), $\varepsilon_{FE}$ ~10. Putting the values in equation (2), effective field across the oxide $E_{DE}$ is V/7 while the field across ferroelectric, $E_{FE}$ is $V_G$/17.5, where $V_G$ is the sweeping voltage magnitude at the gate terminal. This reduced field across the ferroelectric causes a gradual polarization reversal. However, for a significant memory window, the field across the ferroelectric needs to be high enough to switch enough number of dipoles. Since P(VDF-TrFE), has a coercive field of 50 MV/m, driving them to saturation using <4V needs optimized device geometry. Alternately, operation along a hysteresis minor loop can also to be considered.

In FeFET2, a voltage drop across the 3 nm dielectric does not allow enough voltage drop across the 10 nm FE and therefore, partial FE polarization switching leads to gradual change in $I_d$. However, with a channel width, $W = 20$ µm and channel length $L = 10$ µm and a channel area of 200 µm$^2$, we found a rather clear Off to On switching and a MW of nearly 1.5 V when operated within -1 to +3V range (**Figure 3(f)**). For a read bias of 1.5 V, an On/Off ratio of 100 can be achieved. This is significantly higher compared to the FeFETs with $W = 50$ µm and $L = 5$ µm and a channel area of 250 µm$^2$. Area dependence of switching field and timescales have been reported previously in literature for P(VDF-TrFE) based FeFETs. [41, 42] P(VDF-TrFE) is a molecular ferroelectric with several micrometer long polymeric chains. The $CH_2$-$CF_2$ molecular dipoles in the P(VDF-TrFE), responsible for the net electric polarization in the polymer chains have a restricted degree of freedom with main rotation happening only about the chain axis. Since switching of the ferroelectric in our devices involves the rotation of the molecular dipoles about the chain axis, the switching field of the devices become area dependent. With larger device area, more polymeric chains couple with each other restricting the molecular motion through interchain and intrachain interaction. Because of this short and long-range in-plane coupling, the required switching field increases with the gated device area.

Previously, it was shown that through nano-patterning of the polymer ferroelectric, [41] the coercive voltages of P(VDF-TrFE) can be brought down drastically by reducing the switchable polymer chain length. This attribute of the polymeric ferroelectric could serve as a promising route for engineering the needed switching field by modifying the channel area of these devices making them suitable for integration with 180 nm CMOS technologies where maximum driving voltage is <3.6V.

Another important point to consider is the different amount of charge trapping at the $MoS_2$-P(VDF-TrFE) interface for different channel lengths. In a previous work, [43] it has been shown that modified electrostatic coupling of the channel to the drain due to changing $V_d$ can change $C_{FE}$ and $C_{DE}$ which, in turn, can modify the MW and the threshold voltage. In our devices, charge trapping at the $MoS_2$-P(VDF-TrFE) interface could modify the $C_{FE}$ changing the voltage across the FE. Different channel geometries and interface layers need to be studied to have a thorough understanding of this effect.

## 2.2. Programming retention, intermediate states and device endurance

Retention of the programmed states has been measured using a write once read many protocol, as shown in **Figure 4(a)**. In the FeFET1 structure, where the 30 nm P(VDF-TrFE) layer is sandwiched between the $MoS_2$ channel and the gate electrode, without an additional dielectric layer, the measured written states show non-volatile behavior up to the characterization period of 300 s. For the measurements, we used a few (2-3) pulsed shot programming of a certain conductance state, followed by repeated reading at $V_{read}$ = 0.5 V. For the bare P(VDF-TrFE) films, PFM and KPFM measurements (**figure S4**) clearly indicated no significant change in ferroelectric polarization state for the one hour duration of the experiment. The data and measurements shown in supplementary **figure S4** confirm the non-volatile data retention in our ferroelectric only devices. However, for FeFET2, the retention characteristics show a different behavior. For these devices, a distinct decay in the drain current level was observed for all programmed current levels immediately after withdrawal of the $V_{write}$. However, after an initial drop, the read current was stabilized after about 10s and multiple distinguishable current states could be read throughout the measured 1000 s measurement period (**Figure 4(b)**).

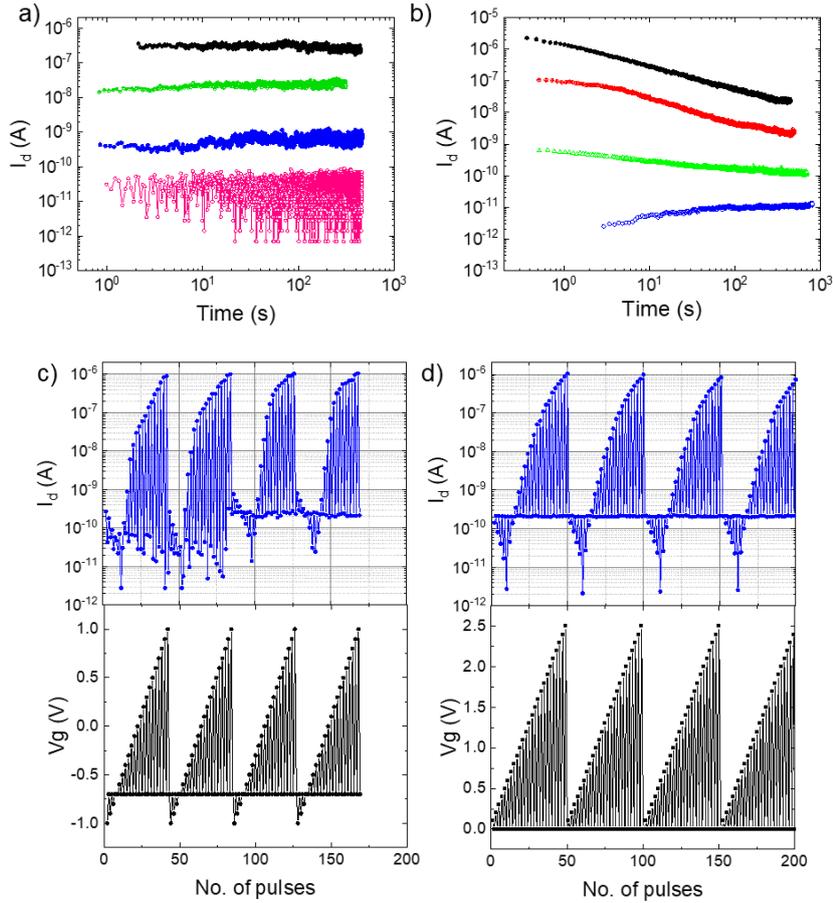

**Figure 4.** (a) Retention properties of different drain currents of (a) FeFET1 and (b) FeFET2 devices. For the measurements, devices were programmed once with a certain applied gate voltage $V_{write}$ followed by repeated reading of the programmed states using $V_{read}$. The results show that in FeFET1, the programmed states are stable over the entire measured duration of 400s while for FeFET2, a distinct decay in the current level was seen immediately after withdrawal of the $V_{write}$. After the initial decay, however, the current stabilized for all programmed states.

Decay in the programmed states of a ferroelectric thin film device may be attributed to the increased depolarizing field ($E_{dep}$) in the ferroelectric due to two main reasons. 1) Incomplete screening of polarization charges in the FE layer due to the presence of an insulating layer at one side of the ferroelectric. 2) Thinner ferroelectric layer causing higher depolarization field leading to faster relaxation of polarization charges. In a heterostructure consisting of alternating FE and DE films (i.e. our FeFET2 architecture), depolarization field can impact in different ways. First, to screen the $E_{dep}$, the FE stack can decompose into a large number of domains. The free energy of the stack has two components, the electrostatic energy ($E_{ES}$) and the domain wall energy ($E_{DW}$). [44] The components have a subtle balance that depends critically on the thickness and the permittivity of the DE and FE, respectively. With decreasing domain width, $E_{ES}$ decreases, however, the energy cost associated with creating the domain walls ($E_{DW}$) increases.

Finally, the domain density is governed by the minimal free energy in the energy landscape for domain reconfiguration. For a thin DE spacer, less screening of $E_{dep}$ is required and hence, the domain density is smaller. For thicker DE spacer, the domain density increases for compensating $E_{dep}$. Grains in polycrystalline films have different crystallographic orientations making the local electric field distribution experienced by the domains quite complex. This causes the domains to have polarization axes oriented in different directions which leads to gradual polarization switching. An optimal balance between the dielectric and ferroelectric thickness and permittivity can ensure enough discrete domains to achieve analog states while avoiding compromise on the data retention time due to increased depolarization field. $E_{dep}$ causes a significant distribution the coercive fields of the domains. This is due to the higher energy barrier needed to be overcome the polarization reversal. As a result, the switching field of the domains and their characteristic switching times are increased. [45,46] Optimising for all the above would yield ferroelectric memories with analog states and good retention.

**2.3. Pulsed weight update and Synaptic plasticity**

Reproducibility of multiple analog states are essential for online training of the synaptic devices, and we tested the reproducibility for both FeFET1 and 2 devices over multiple cycles using an increasing pulsed programming scheme (**Figure 4**(**c** and **d**)). The results demonstrate high endurance over ten thousand switching cycles for both kind of FeFETs, making these devices suitable, not only for non-volatile data retention, but as analog memories and electronic synaptic transistors. However, compared to FeFET1, intermediate states were more reproducible and controllable in FeFET2, as expected from their respective switching dynamics. Here, it is important to mention that under the employed pulsing schemes, both these devices are operated in their minor field loops with the ferroelectric polarization switched only partially.

The conductance of both FeFET1 and FeFET2 were recorded using single-shot gate voltage pulses of constantly increasing magnitude to achieve multiple conductance states. Before applying each programming pulse, application of an erase pulse ensured the device gets back to its Off state. **Figure 4(c and d)** shows multiple intermediate conductance states ranging from 100 pA to 1 µA with the intermediate states being more controllable for FeFET2 compared to FeFET1. For FeFET1, the abrupt switching during the Off to On transition, arising from sharp ferroelectric polarization switching, results in nonlinear conductance modulation in the pA to tens of nA. For FeFET2, the reproducibility of the conductance linearity was easier to obtain due to gradual polarization switching arising from the reduced voltage drop over the ferroelectric layer in these devices. The current range between 20 nA to 1.2 µA was chosen for testing the synaptic computational capability of these devices and variation of individual programmed states from cycle to cycle were analyzed for FeFET2 (discussed later, during the training operation). Drain current below tens of nanoamperes can result in higher read noise and slower readout and therefore avoided for training experiments.

**2.4. DNN Training**

Previously, it has been shown that monolithic crossbar or pseudo crossbar arrays with analog non-volatile memories can store and update weights on-chip, offering the possibility of accelerating DNN training. However, with analog memories, the reported training accuracy of DNNs is often below the required level due to device non-idealities, like nonlinearity, asymmetry, limited bit precision, and limited dynamic range of the weight updates. Additionally, limited endurance of these devices degrades the training accuracy even further.

We studied the gate voltage-controlled dynamics of partial polarization switching in our $MoS_2$ based FeFETs to demonstrate efficient DNN training by utilizing the analog weight update in FeFET-based synapses. The training simulation was performed on MNIST handwritten dataset of small and large dimensions (64×36×10 and 784×300×10, respectively). During each epoch, the network was trained on 60 000-image training data set, and the recognition accuracy was tested on a separate 10 000-image testing data set. The simulated network consists of one input layer, one hidden layer and one output layer of neurons (**Fig. 5(a)**).

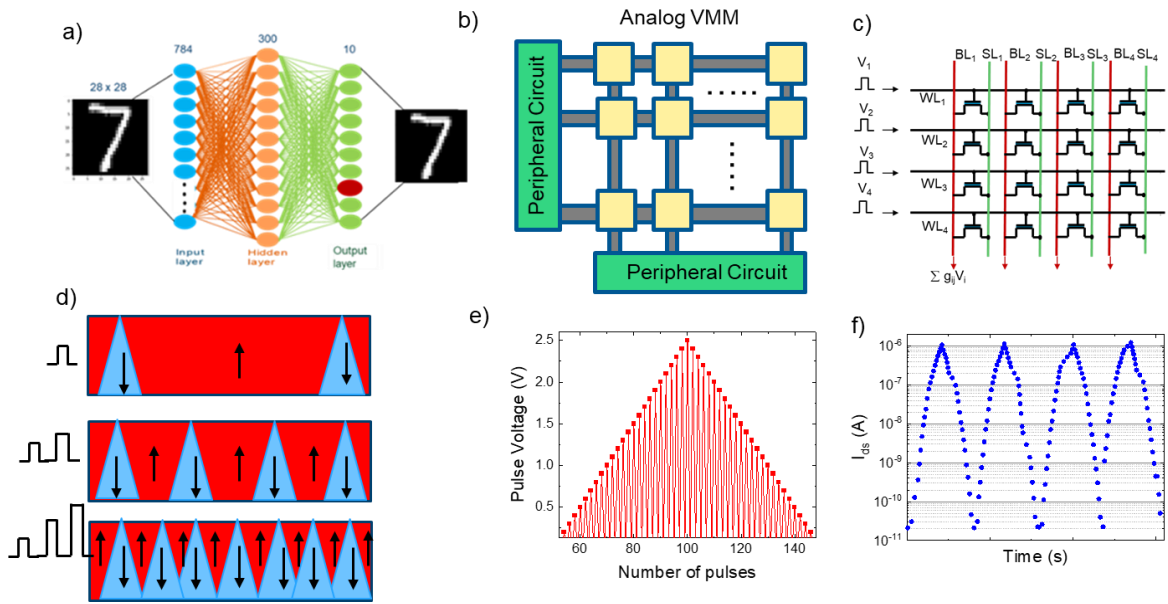

**Figure 5.** (a) MNIST handwritten digit recognition using a 3-layer DNN containing one input, one output and one hidden layer. (b) Schematic diagram of the vector matrix multiplication (VMM) architecture using FeFET based crossbar arrays where (c) voltage pulses are fed to the gate terminals (WLs) and current outputs are extracted from the drain current terminals (BLs).

(d) Schematic representation of partial polarization switching in ferroelectric gate due to application of voltage pulses of increasing amplitude. (e) Actual voltage pulses applied to FeFET2 for one single potentiation and depression cycle in our experiments and (f) corresponding channel currents in FeFET2 devices over multiple switching cycles.

**Figure 5(b)** shows the schematic diagram of the vector matrix multiplication (VMM) operation done in FeFET based analog crossbar arrays where each cross point between row and columns represents one FeFET2 transistor. A single FeFET array core, where the VMM operation takes place, is surrounded by neuron circuitry which implements forward propagation, back propagation, and weight-update operations. For maximizing the performance and parallelism, each row and column has its own neuron circuitry that helps to avoid time-multiplexing. **Figure 5(c)** shows experimental protocol of feeding the voltage pulses to the gate terminals (WLs) of the memory transistors where the current outputs are extracted from the drain current terminals (BLs) during the training operation in simulation. **Figure 5(d)** shows the schematic representation of partial polarization switching in ferroelectric gate insulator due to the application of voltage pulses of increasing amplitude. This partial polarization switching changes the channel conductance gradually leading to an analog synaptic behavior. The actual voltage pulses and the corresponding channel currents in FeFET2 devices are shown in **Figure 5 (e)** and **(f)**, respectively showing nearly symmetric potentiation and depression cycles of post-synaptic currents with good reproducibility.

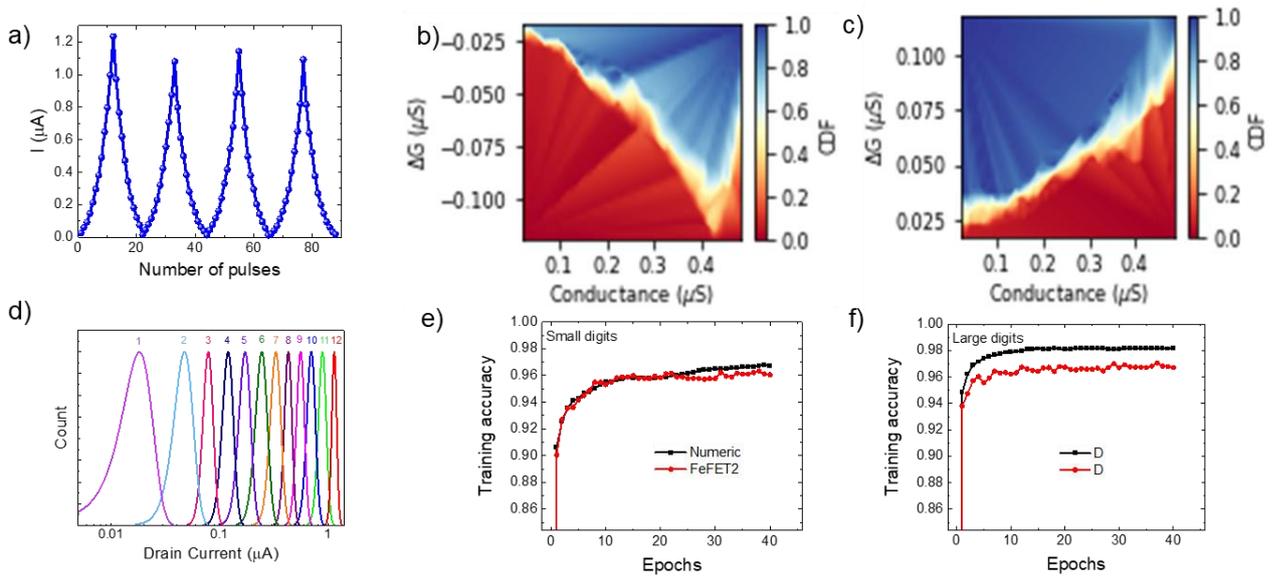

**Figure 6.** (a) Potentiation and depression cycles of FeFET2 in response to increasing and decreasing set of programming and erasing voltage pulses showing large number of reproducible intermediate conducting states achievable with satisfactory level of linearity. Distribution of $\Delta G$ as a function of $G$ for typical (b) potentiation and (c) depression cycles and the cumulative distribution function of $\Delta G$ in each cycle showing the conductance variability

that the training operation needs to accommodate. (d) Distribution of the chosen 12 conductance states that are used for the training simulation over multiple cycles of weight update. Training accuracy of the FeFET based DNNs on (e) small MNIST dataset and (f) large MNIST dataset.

**Fig. 5(f)** shows the entire dynamic conductance range of our FeFET2 devices. For efficient DNN training, synaptic weight elements are expected to satisfy certain conditions that include a symmetric and linear conductance response with ≥32 conductance states (≥5-bit), and a maximum $G$ ($G_{max}$)/minimum $G$ ($G_{min}$) ratio of >10. [5] In our previous work, we have shown that it is challenging to obtain large dynamic conductance range ($G_{max}/G_{min}$) in FTJ devices with NLS domain switching characteristics, with programming pulses of 1 V magnitude and ns durations. [14] In comparison, FeFETs show much higher $G_{max}/G_{min}$ ratio due to the large On/Off ratio of the FETs. However, the entire dynamic range cannot be used for training in most cases due to lack of linearity and symmetry in response to potentiating and depressing pulses or due to the challenge of reading out too low current due to interfering read noise. In our FeFET architectures under study, the conductance update linearity is improved significantly in FeFET2 compared to FeFET1 due to gradual nature of ferroelectric switching. The neural network training simulation was performed on a FeFET2 based neural network architecture to test our device performance on classification of standard MNIST handwritten dataset. [47]

To evaluate the effect of non-ideal FeFET conductance levels and noise on the accuracy of neural training algorithm, we performed simulations using a modified version of Sandia CrossSim simulator on the standard MNIST handwritten digit dataset. [48] The neural network code is based on backpropagation (BP) algorithm, and we used our measured FeFET2 potentiation and depression characteristics over multiple switching cycles for the simulation. BP is one of the most numerically demanding and energy inefficient neuromorphic algorithms used in a CMOS architecture. Therefore, running the BP algorithm on a FeFET based analog neural architecture provide a pathway for performance acceleration in terms of energy and computational efficiency. [5] In our simulation, the FeFET based crossbars are used as part of a neural core, i.e., a single layer of the DNN that accelerate the main matrix operations while a digital core is used to process the inputs and outputs to the crossbar. The conductance values change slightly from cycle to cycle, as shown in **Figure 6(a)**. For our training experiments, the drain current between 25 nA to 1.2 µA was chosen. To measure the weight update linearity under continuous pulse trains, gradually increasing or decreasing pulsing schemes of duration 100 ms were used. Before training the devices were programmed to full Off state and then, alternating potentiating and depressing pulse trains were applied.

Precise control of the ferroelectric domain dynamics is an essential criterion for linear weight update in ferroelectric synaptic devices and a proper gate stack design in FeFETs is critically important for obtaining this linearity. As discussed previously, we found that in FeFET2, it is possible to have more controlled conductance modulation through partial rotation of polarization. The distribution of $\Delta G$ as a function of conductance $G$ for a typical depression and potentiation cycles are shown in **Figure 6(b) and (c)** respectively and the cumulative distribution function (CDF) of $\Delta G$ for multiple cycles is shown in the right axes. The randomness of domain switching, an inherent property arising from nucleation limited switching in polycrystalline FE films, can explain the noise margin of the conductance states over cycle to cycle. This could potentially affect the training accuracy of the circuits. Therefore, in our experiments, the network training operation has been performed by utilizing weight update data over several tens of potentiation and depression cycles, minimizing the effect of read noise on the training accuracies. In **Figure 6(d)**, the standard distribution of 12 different conductance states, chosen for training is shown.

In our simulation, the BP algorithm has been implemented utilizing the following procedure. Each layer of the DNN is represented by a simulated crossbar of FeFETs that performs two key operations, a VMM and a parallel outer product update. All FeFET devices in the crossbar array are updated in parallel, leading to efficient one-step weight updates that improves the energy efficiency and latency of the system. The VMM operation in the analog core takes place by utilizing Ohm's law and Kirchoff's current law. The voltage pulses are sent to the gate terminal rows (horizontal lines) of the crossbar and the current output is extracted from the drain terminal column (vertical lines) of the crossbar. A sigmoid neuron with a slope of 1 and a learning rate of 0.1 is used to perform classification tasks with a high level of accuracy. In BP, the actual outputs of the computing circuits are compared to the expected output after each training cycle and the difference (i.e., the error) is propagated back to the circuit for making a correction on the weights, $G$.

In **Figure 6(e) and (f)**, the numerical and actual hardware-based training accuracies are compared. Where a standard double-precision CPU or GPU can provide a training accuracy of 98%, a FeFET2 based analog accelerator could provide nearly 96% accuracy for both large and small digits. Number of parameters involved in computations in neural networks scale up drastically with the size of the network. This imposes a huge burden on hardware resources, restricting the processing speed, and consuming huge amount of energy. In our training simulation, we used both large and small handwritten digit recognition to clarify the effect of two different sizes of the network on the device non-idealities like non-linearity and asymmetry.

As can be seen from **Fig. 6(e) and (f)**, the difference between numerical training accuracy and experimental training accuracy starts to grow significantly as the size of the network increases. By modifying pulse widths, synaptic training conductance range or circuit parameters like A/D or D/A conversion quantization, it is possible to obtain improved training accuracy in larger networks.

**Benchmarking synaptic behavior and training accuracy**

Finally, we benchmark our FeFET based DNN accelerator performance with other state-of-the art analog DNN accelerator implementations. Here, we focus on the ferroelectric NVM based neural network training, where multi-layer perceptron (MLP) based network training simulation is performed on MNIST handwritten dataset. It is evident from the comparison that our $MoS_2$ based FeFET based classifier performance is at par with the best reported results. Application of incremental pulsing schemes for weight update linearity is still an open issue that needs to be addressed in our future works.

**Table 1.** Benchmarking ferroelectric memory based DNN accelerator performance.

| FE Material | Technology | $G_{max}/G_{min}$ | Pulse Scheme | Maximum $V$ | $R_{On}$ ($\Omega$) | Number of states | Training accuracy | Ref. |
|---|---|---|---|---|---|---|---|---|
| HZO | FeFET | 45 | Incremental | 3.65V, 75 ns | 500 k | 32 | 90% | 5 |
| HZO | FeFinFET | 45 | Incremental | 8V, 100 ns | 10 M | 27 | 97% | 49 |
| HZO | FTJ | 2 | Identical | 3V, 100 ns | 10 k | 10 | 96% | 50 |
| HZO | FeTFT | 100 | Incremental | 6V, 100 ns | 100 k | 8 | 95% | 51 |
| P(VDF-TrFE) | FTJ | 1.2 | Identical | 0.8V, 100 ms | 10 k | > 8 | 93-95% | 14 |
| P(VDF-TrFE) | FeFET **This work** | >50 | Incremental | 2.5V, 100 ms | 1 M | > 16 | 96% | - |

In previous works on flexible FeFET based neural networks, the influence of the FE properties and contact barrier heights in the P(VDF-TrFE) based FeFET synapses was demonstrated.[52] Performance was evaluated on the training and recognition performance of a multi-layer perceptron (MLP)-based neural network. Higher annealing temperature of P(VDF-TrFE) and the nature of the contact metals (Ti, Cr, Pd), providing a relatively high tunneling injection barrier of the FeFET, was shown to improve the dynamic conductance range and nonlinearity of the post synaptic current. This improvement in dynamic range and nonlinearity result in improved accuracy of recognition in neural networks. The effect of the contact metal on the recognition rate revealed that the synapses with Pd-contact can achieve maximum recognition rate compared to Ti or Cr-contacted synapses. This result opens the prospect of future optimizations of P(VDF-TrFE) morphology, semiconductor-FE interface and contact metals for improving cognitive capabilities of P(VDF-TrFE) based FeFETs. In another work, stable synaptic functionalities under extreme bending conditions of 50 µm radius were demonstrated in ultrathin conformable organic synaptic transistors based on P(VDF-TrFE). [53] These results, together with our FeFET based neural network performance, opens a future direction of research for high-performance intelligent wearables, operating at affordable power budgets.

**Conclusion**

In conclusion, we report a CMOS BEOL and flexible substrate compatible analog synaptic transistor with more than 16 distinguishable analog states that can work as a building block for deep neural network accelerators. The building block, a FeFET based on ferroelectric polymer gate insulator P(VDF-TrFE) and 2D semiconductor $MoS_2$ as the *n*-type semiconducting channel material can work both as digital and analog memory based on the device gate stack design. The analog conductance of the FeFETs is achieved by employing a hybrid ferroelectric-dielectric layer at the gate stack. The ferroelectric only FeFETs with 30 nm P(VDF-TrFE) layer, show promising features as a digital memory with the OFF/ON ratio of $\approx 10^6$, memory window of 3V operating at $<\pm 5V$ and a long data retention time. The hybrid ferroelectric-dielectric FeFETs, with 10 nm P(VDF-TrFE) layer and 3 nm $Al_2O_3$ show quasicontinuous resistive switching in the $MoS_2$ channel resulting from the gradual ferroelectric domain switching. Under increasing and decreasing pulsed voltage schemes, the FeFETs produced conductivity change of $10^4$ with multiple reproducible states. Network training experiments performed with the analog FeFETs within the current range of 20 nA to 1.2 µA show 96% classification accuracy for the MNIST handwritten datasets, underlining their immense potential for online learning and implementation in flexible and wearable cognitive devices.

**Experimental Methods**

*Device Fabrication*: 2-3 layers of $MoS_2$ were deposited on an ≈ 285 nm thick $SiO_2$ dielectric layer on top of a (100) p-type Si wafer (resistivity 1-10 Ω cm) using the procedure described in our previous work. [37] The $MoS_2$ films were then transferred to a 30 nm thick ALD grown

Al$_2$O$_3$ layer on polyimide systrate on Si backplate. After transfer, the MoS$_2$ channel area was defined using a mask-less laser writer lithography process and the rest of the MoS$_2$ was etched using SF$_6$ and O$_2$ plasma. Drain-source contacts were patterned on top of MoS$_2$ using a lithography step and Ti (2 nm) and Au (50 nm) electrodes were deposited by e-beam evaporation at room temperature. The device was then annealed at 200 °C in a vacuum furnace for 2 h to remove resist residues and to improve the contact resistance. For the P(VDF-TrFE) (70:30 mol%) copolymer film, ferroelectric polymer powder from Piezotec was dissolved in the Methyl Ethyl Ketone (MEK) with 0.5 or 0.2 wt% strength and the solution was spin coated on top of MoS$_2$. First, the spin-coated films were dried at 100 °C for 10 minutes and then the films were annealed at 135 °C for 2 h at vacuum for improving their crystallinity. Transistor devices with metal-ferroelectric-semiconductor (MFS) structure in the gate stack, referred as FeFET1, consists of a 30 nm thick P(VDF-TrFE) gate insulator layer directly between the MoS$_2$ channel and gate metal electrode. Transistors with metal-insulator-ferroelectric-semiconductor (MIFS) structure in the gate stack, referred as FeFET2, consists of a 10 nm P(VDF-TrFE) layer together with 3 nm Al$_2$O$_3$ layer, deposited by ALD at 120 °C, embedded between the ferroelectric and the gate metal. Finally, 30 nm thin Au films were deposited by e-beam evaporation and patterned by lithography as the top gate electrode. The thickness of the ferroelectric films was characterized from a Si/SiO$_x$/P(VDF-TrFE)/Au capacitor structure using profilometry and AFM. The memory window scaling study was performed on ITO/ P(VDF-TrFE)/Au capacitors using an impedance analyzer. The process flow for device fabrication is given in Fig. *S1* of supplementary information.

*Structural Characterization:* The ALD grown MoS$_2$ films and spin-coated P(VDF-TrFE) films were characterized by atomic force microscopy (AFM), micro-Raman measurements and local piezo-force microscopy measurements. For all the above-mentioned characterization, a thin film of either MoS$_2$ or P(VDF-TrFE) layer was used on a conducting substrate.

*Raman Characterization:* Raman spectroscopy of the MoS$_2$ films were done with a Renishaw Invia system using a 532 nm laser and a x50 objective.

*Piezo-force Microscope (PFM) Characterization:* Atomic Force Microscopy (AFM), Kelvin-probe force microscopy (KPFM) and local PFM measurements of typical 30 nm P(VDF-TrFE) film on Indium Tin Oxide (ITO) substrate was performed in a Bruker Dimension 5000 system in tapping mode using Pt/Ir-coated Si cantilever tips by superimposing an ac voltage of 1 V amplitude onto the sweeping dc bias voltage. The voltages were applied to the cantilever tip keeping the bottom ITO electrode grounded. The PFM system was operated near a resonance frequency of 350 kHz. For KPFM measurements, the ferroelectric films were first poled using the AFM tip in contact mode followed by non-contact probing of the surface potential of the films due to the bound polarization charges on the ferroelectric measured over a period of 1 hour.

*Electrical Measurements:* All electrical measurements were performed using a Keithley 4200A-SCS parameter analyzer under room temperature and ambient conditions. Continuous transfer characteristics measurements were performed by applying sweeping voltage at the top gate under a continuous bias voltage between drain and source contacts. For pulsed measurements, pulses of varying magnitude were applied at the top gate. The constant width of all voltage pulses was 100 ms. All measurements were carried out under ambient conditions without illumination.

**Supporting Information**

Supporting Information is available from the Wiley Online Library.

**Acknowledgements**

SM thanks the Academy of Finland (Grant no. 345068 and 350667) for financial support. The work used experimental facilities of Micronova National Research Infrastructure for Micro- and Nanotechnology.

**References**

1. D. Ciregan, U. Meier and J. Schmidhuber, *2012 IEEE Conference on Computer Vision and Pattern Recognition* **2012**, 3642-3649.
2. G. Hinton, L. Deng, D. Yu, G. Dahl, A.-r. Mohamed, N. Jaitly, A. Senior, V. Vanhoucke, P. Nguyen, T. Sainath, B. Kingsbury, https://static.googleusercontent.com/media/research.google.com/en//pubs/archive/38131.pdf.
3. S. Park, A. Sheri, J. Kim, J. Noh, J. Jang, M. Jeon, B. Lee, B. R. Lee, B.H. Lee, H. Hwang, *2013 IEEE International Electron Devices Meeting*, **2013**, pp. 25.6.1-25.6.4.
4. T. Gokmen, Y. Vlasov, *Front. Neurosci.* **2016**, *10*, 1-13.
5. M. Jerry, P.-Y. Chen, J. Zhang, P. Sharma, K. Ni, S. Yu, S. Datta, *2017 IEEE International Electron Devices Meeting (IEDM)*, **2017**, 6.2.1-6.2.4.
6. G. W. Burr et al., *Adv. Physics: X* **2017,** *2*, 89–124.
7. D. Ielmini, S. Ambrogio, *Nanotechnology* **2020**, *31*, 092001.
8. J. Woo, S. Yu, *IEEE Nanotech. Magaz.*, **2018**, *12*, 36-44.
9. I. Boybat et al., *Nat. Commun.* **2018**, *9*, 2514; M. Ito et al., *2018 IEEE 18th International Conference on Nanotechnology* (*IEEE-NANO*), **2018**, 1-4.
10. A. I. Khan, A. Keshavarzi, S. Datta, *Nat. Electron.* **2020**, *3*, 588–597.
11. S. Oh, H. Hwang, I. K. Yoo, *APL Materials* **2019**, *7*, 091109.
12. S. Majumdar, H. Tan, Q. H. Qin, S. van Dijken, *Adv. Electron. Mater.* **2019**, *5*, 1800795.
13. S. Majumdar, *Nanoscale* **2021**, *13*, 11270.
14. S. Majumdar, *Neuromorphic Comp. Eng.* **2022**, *2*, 041001.
15. J. Y. Kim, M.-J. Choi, H. W. Jang, *APL Materials* **2021**, *9*, 021102.
16. X. Li and H. Zhu, *J. Materiomics*, **2015**, *1*, 33-44.
17. L. T. Liu, S. B. Kumar, Y. Ouyang, and J. Guo, *IEEE Trans. Electron Dev.*, **2011**, *58*, 3042-3047.
18. S. Majumdar, *Adv. Intell. Syst.* **2022**, *4*, 2100175.
19. S. Majumdar, H. Tan, I. Pande, S. van Dijken, *APL Mater.* **2019**, *7*, 091114.
20. M. Fitsilis, https://publications.rwth-aachen.de/record/62096/files/Fitsilis_Michael.pdf.
21. J. Li, M. Si, Y. Qu, X. Lyu, P. D. Ye, *IEEE Trans. Electron. Dev.* **2021**, *68*, 1214-1220.
22. K. A. Aabrar et al., *IEEE Trans. Electron. Dev*. **2022**, *69,* 2094-2100.
23. Y. Sun, G. Niu, W. Ren, J. Zhao, Y. Wang, H. Wu, L. Jiang, L. Dai, Y.-H. Xie, P. R. Romeo, J. Bouaziz, B. Vilquin, *AIP Advances* **2021**, *11*, 065229.
24. S. Horiuchi, Y. Tokura, *Nat. Mater.* **2008**, 7, 357–366.
25. X. Chen, X. Han, Q.-D. Shen, *Adv. Electron. Mater.* 3, 1600460, 2017.
26. T. Xu, L. Xiang, M. Xu, W. Xie, W. Wang, *Sci. Rep.* **2017**, *7*, 8890.
27. X. Wang, Y. Chen, G. Wu, D. Li, L. Tu, S. Sun, H. Shen, T. Lin, Y. Xiao, M. Tang, W. Hu, L. Liao, P. Zhou, J. Sun, X. Meng, J. Chu, J. Wang, *npj 2D Mater. Appl.* **2017**, 1, 38.
28. B. Tian, L. Liu, M. Yan, J. Wang, Q. Zhao, N. Zhong, P. Xiang, L. Sun, H. Peng, H. Shen, T. Lin, B. Dkhil, X. Meng, Chu, X. Tang, C. Duan, *Adv. Elec. Mater.* **2019**, *5*, 1800600.
29. S. Kim, K. Heo, S. Lee, S. Seo, H. Kim, J. Cho, H. Lee, K.-B. Lee, J. -H. Park, *Nanoscale Horiz.* **2021**, *6*, 139.
30. A. Van Breemen, B. Kam, B.Cobb, F. Gonzales Rodriguez, G. van Heck, K. Myny, A. Marrani, V. Vinciguerra, G. Gelinck, *Org. Electron.* **2013**, *14*, 1966–1971.

31. B. B. Tian, J. L. Wang, S. Fusil, Y. Liu, X. L. Zhao, S. Sun, H. Shen, T. Lin, J. L. Sun, C. G. Duan, M. Bibes, A. Barthélémy, B. Dkhil, V. Garcia, X. J. Meng, J. H. Chu, *Nat. Commun.* **2016**, 7, 11502.
32. S. Majumdar, B. Chen, Q. H. Qin, H. S. Majumdar, S. van Dijken, *Adv. Func. Mater.* 2018, 28, 1703273.
33. T. Xu, L. Xiang, M. Xu, W. Xie, W. Wang, *Sci. Rep.* **201**, *7*, 8890.
34. https://irds.ieee.org/editions/2022
35. H. S. Lee et al., *Small* **2012**, *8*, 3111–3115.
36. P.-C. Shen, C. Lin, H. Wang, K. H. Teo, J. Kong, *Appl. Phys. Lett.* **2020,** *116*, 033501.
37. N. Aspiotis, K. Morgan, B. März, K. Müller-Caspary, M. Ebert, C. C. Huang, D. W. Hewak, S. Majumdar, I. Zeimpekis, *npj 2D Mater. Appl.* (in press, 2023) *arXiv preprint arXiv:2203.10309*.
38. J. J. Brondijk, K. Asadi, P. W. M. Blom, D. M. de Leeuw, *J. Polym. Sci., Polym. Phys* **2012***, 50*, 47-54.
39. V. Georgiou, D. Veksler, J. P. Campbell, P. R. Shrestha, J. T. Ryan, D. E. Ioannou, K. P. Cheung, *Adv. Func. Mater.* **2018,** *28*, 1705250.
40. M.D. Groner, J.W. Elam, F.H. Fabreguette, S.M. George*, Thin Solid Films* **2002,** *413*, 186–197.
41. Z. Hu, M. Tian, B. Nysten, A.M. Jonas, *Nat. Mater.* **2008**, *8*, 62; A. Jonas, Z. Hu, WO 2009/144310 Al, **2009**.
42. S. Das, J. Appenzeller, *Organic Electronics* **2012,** *13*, 3326-3332.
43. X. Yin, A. Aziz, J. Nahas, S. Datta, S. Gupta, M. Niemier, X. S. Hu, *2016 IEEE/ACM International Conference on Computer-Aided Design (ICCAD)* **2016**, 1-8.
44. A. Kopal, P. Mokrý, J. Fousek and T. Bahník, *Ferroelectrics* **1999**, *223*, 127-134, .
45. D. Zhao et al., *Nat. Commun.* **2019**, *10*, 1-11.
46. M. Mai, B. Martin and H. Kliem, *J. Appl. Phys.* **2011,** *110*, 064101.
47. Y. Le Cun, C. Cortes, C. J. Burges, The MNIST database of handwritten digits [online] Available: http://yann.lecun.com/exdb/mnist
48. S. Agarwal et al. *2016 International Joint Conference on Neural Networks (IJCNN)* **2016**, pp. 929-938
49. S. De et al., *2021 Symposium on VLSI Technology* **2021** pp. 1-2.
50. L. Chen et al., *Nanoscale* **2018**, *10*, 15826 - 15833.
51. S. De et al., *TechRxiv. Preprint.* https://doi.org/10.36227/techrxiv.19491221.v2
52. S. Kim, K. Heo, S. Lee, S. Seo, H. Kim, J. Cho, H. Lee, K.-B. Lee, J.-H. Park, *Nanoscale Horiz.*, **2021**,*6*, 139.
53. B. Tian, L. Liu, M. Yan, J. Wang, Q. Zhao, N. Zhong, P. Xiang, L. Sun, H. Peng, H. Shen, T. Lin, B. Dkhil, X. Meng, J. Chu, X. Tang, C. Duan, *Adv. Electron. Mater.* **2019**, *5*, 1800600.

The table of contents entry should be 50–60 words long and should be written in the present tense. The text should be different from the abstract text.

C. Author 2, D. E. F. Author 3, A. B. Corresponding Author* ((same order as byline))

**Title** ((no stars))

ToC figure ((Please choose one size: 55 mm broad × 50 mm high **or** 110 mm broad × 20 mm high. Please do not use any other dimensions))

Supporting Information

**Back-end and Flexible Substrate Compatible Analog Ferroelectric Field Effect Transistors for Accurate Online Training in Deep Neural Network Accelerators**

*Sayani Majumdar\* and Ioannis Zeimpekis*

Sayani Majumdar

VTT Technical Research Centre of Finland Ltd., P.O. Box 1000, FI‐02044, VTT, Espoo, Finland

E-mail: sayani.majumdar@vtt.fi

Ioannis Zeimpekis

Zepler Institute, Faculty of Engineering and Physical Sciences, University of Southampton, United Kingdom

**S1. Device process flow for FeFET1 and FeFET2**

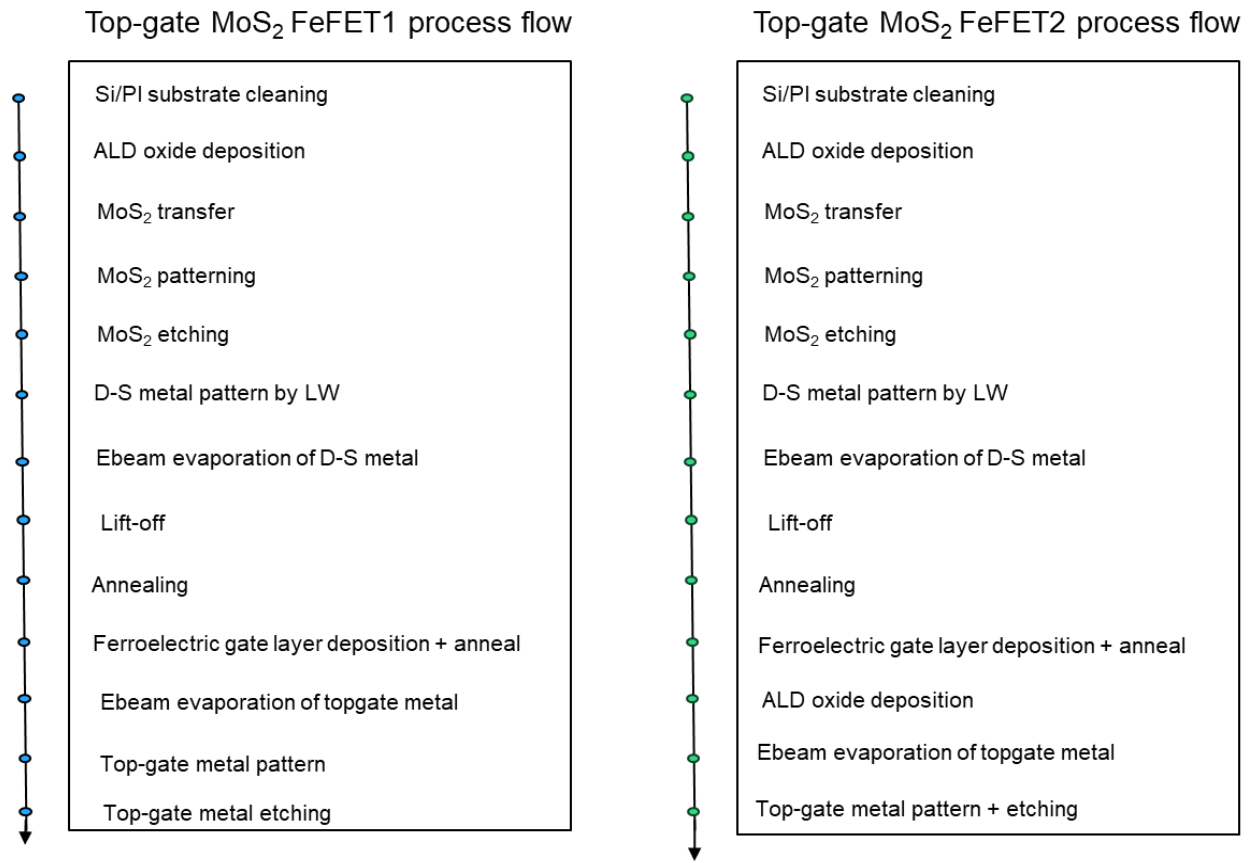

**Figure S1.** The device fabrication process steps for the FeFET1 (left) and FeFET2 (right) architectures.

## S2. MoS₂ transfer process flow for FeFETs

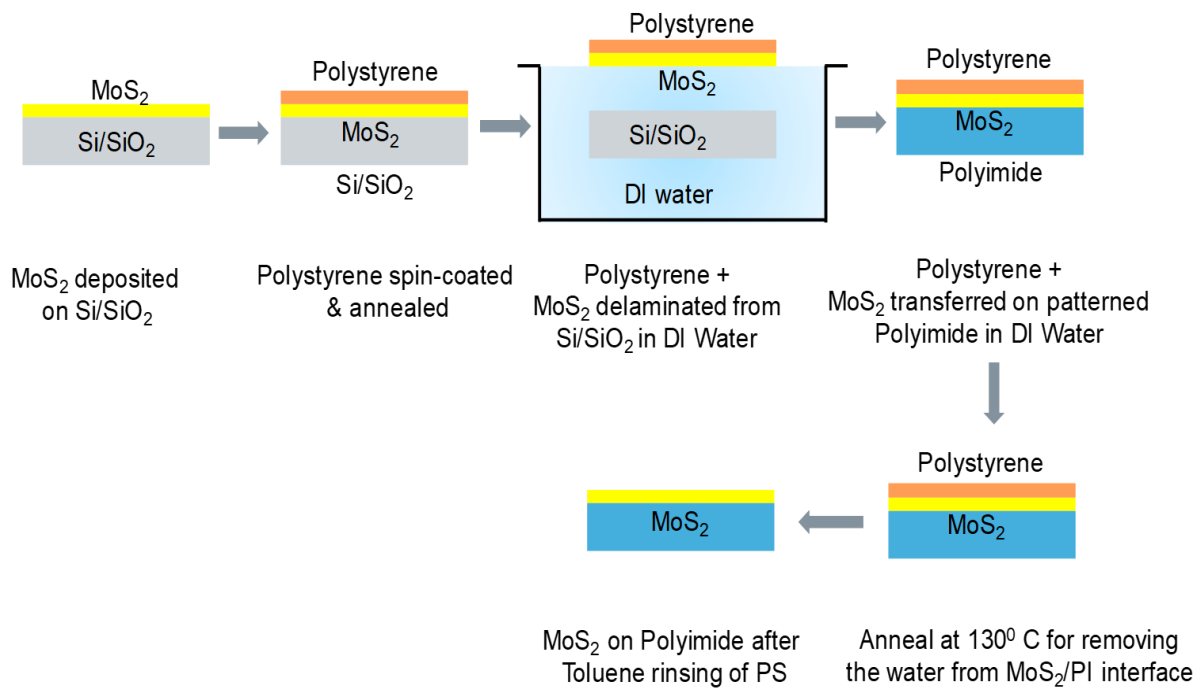

**Figure S2.** Process flow for transferring the MoS$_2$ films on the target substrates. The target substrate in the present work is polyimide (spin coated) on silicon backplate.

**S3. Capacitance measurements of the Metal-Ferroelectric-Metal (MFM) Capacitors**

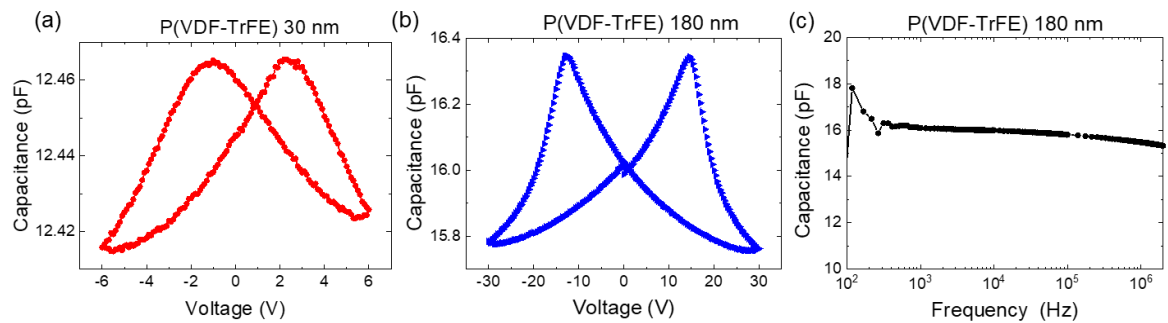

**Figure S3.** (a) Capacitance (*C*) - Voltage (*V*) data of ITO/P(VDF-TrFE)/Au capacitors for (a) 30 nm and (b) 180 nm thicknesses of the ferroelectric layer taken at 1 MHz frequency at 100 mV AC voltage. (c) Capacitance of the 180 nm P(VDF-TrFE) layer as a function of frequency showing upto 2 MHz frequency there is no significant decrease in the capacitance value, emphasizing the possibility of high frequency applications of these devices.

**S4. Non-volatile retention of polarization**

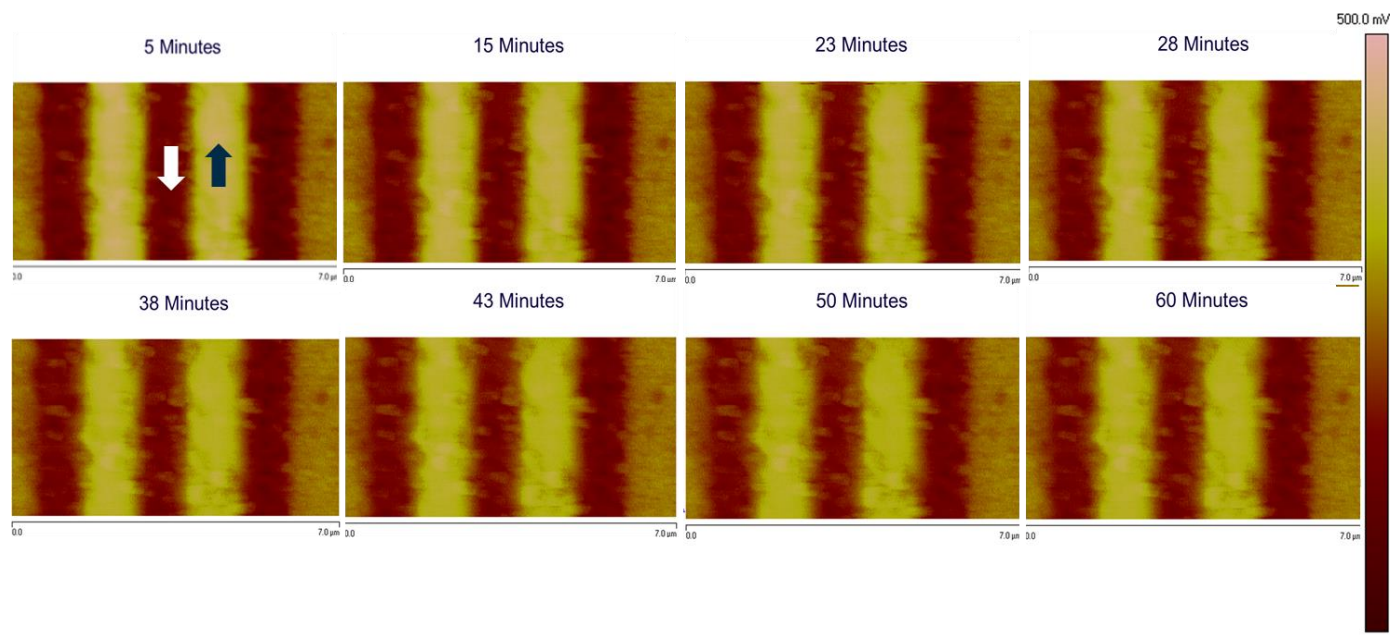

**Figure S4.** Polarization retention study of the P(VDF-TrFE) copolymer thin films. First the samples were poled in positively and negatively poled regions (with the polarization directions indicated by the direction of the arrows in the first image) using the local PFM tip and then Kelvin-probe force microscopy (KPFM) was performed over extended period of time to study the evolution of surface charges over time. The KPFM image of surface potential of P(VDF-TrFE) thin films after poling the domains with applied voltages of ±5V show (From top left to top right: surface potential immediately after poling to first 28 minutes and from bottom left to bottom right: from 38 minutes to 1 hour with multiple measurements in between) long-term stability of the poled domains. The color bar on the right indicates the magnitude of surface potential.